\documentclass[preprint2]{aastex}

\shorttitle{Measuring Be depletion in cool stars with exoplanets}
\shortauthors{Delgado Mena  et al.}

\begin{document}


\title{Measuring Be depletion in cool stars with exoplanets\thanks{Based on observations made with UVES at VLT Kueyen 8.2 m telescope at the European Southern Observatory (Cerro Paranal, Chile) in programme 75.D-0453A}}


\author{E. Delgado Mena\altaffilmark{1,2}, G. Israelian\altaffilmark{1,2}, J. I. Gonz\'alez Hern\'andez\altaffilmark{1,2}, N. C. Santos\altaffilmark{3,4} and R. Rebolo\altaffilmark{1,2,5}}

\altaffiltext{1}{Instituto de Astrofisica de Canarias, 38200 La Laguna, Tenerife, Spain: edm@iac.es}
\altaffiltext{2}{Departamento de Astrof\'isica, Universidad de La Laguna, 38206 La Laguna, Tenerife, Spain}
\altaffiltext{3}{Centro de Astrof\'isica, Universidade do Porto, Rua das Estrelas, 4150-762 Porto, Portugal}
\altaffiltext{4}{Departamento de F\'isica e Astronomia, Faculdade de Ci\^encias, Universidade do Porto, Portugal}
\altaffiltext{5}{Consejo Superior de Investigaciones Cient\'{\i}ficas, Spain}


\begin{abstract}
We present new UVES spectra of a sample of 14 mostly cool unevolved stars with planetary companions with the aim of studying possible differences in Be abundance with respect to stars without detected planets. We determine Be abundances for these stars that show an increase in Be depletion as we move to lower temperatures. We carry out a differential analysis  of spectra of analog stars with and without planets to establish a possible difference in Be content. While for hot stars no measurable difference is found in Be, for the only cool (T$_{\rm eff}\sim$ 5000 K) planet host star with several analogs in the sample we find enhanced Be depletion by 0.25 dex. This is a first indication that the extra-depletion of Li in solar-type stars with planets may also happen for Be, but shifted towards lower temperatures (T$_{\rm eff}$ $<$ 5500 K) due to the depth of the convective envelopes. The processes that take place in the formation of planetary systems may affect the mixing of material inside their host stars and hence the abundances of light elements.
\end{abstract}





\section{Introduction}
The discovery of more than 450 planets in the last fifteen years has given rise to a very successful field in astrophysics. Apart from the study of those planets, the observation of their host stars can provide us with important information about the composition and formation of extrasolar planetary systems.\\

The first abundance analysis of those systems revealed that planet-host stars present considerably higher metallities than single field stars \citep{Gonzalez98, Gonzalez01, Santos00, Santos01, santos04, Fischer}. Moreover, this is not the only difference between stars with and without detected planets. \citet{israelian04} suggested that Li was severely depleted in solar-type stars (stars with T$_{\rm eff}$ between 5600 K and 5850 K) with planets when compared with similar stars without detected planets. This result was latter found by several groups \citep{chen,gonzalez08,gonzalez10,takeda}. \citet{israelian09} confirmed this result using the homogeneous and high quality HARPS GTO sample. Furthermore, stellar mass and age are not responsible for the observed correlation \citep{sousa10}. Indeed, several studies of Li abundances in old clusters like M67 have shown that solar-type stars present a wide dispersion of Li abundances though they have similar ages and metallicities \citep{Randich,Pasquini}. On the other hand, a recent paper by \citet{Baumann} claims that planet-host stars have not experienced an extra Li depletion due to the presence of planets, but due to the higher age of planet hosts. We note that uncertainties in the age determination of main sequence solar-type stars are very large and it is a matter of concern that different methods lead to significant differences in age. Moreover, figure 3 of \citet{Baumann} showing the influence of initial angular momentum on Li depletion is perfectly consistent with the conclusion of \citet{israelian09} and \citet{sousa10}, since the presence of planets could be directly related to the different initial angular momentum \citep{Bouvier}.\\

Several possibilities have been proposed to explain this difference in Li content between stars with and without planets. Pollution, for instance, should be ruled out because it would show, in principle, just the opposite effect, i.e. an increase of the Li abundance in planet-host stars. On the other hand it seems that stars with planets might have a different evolution. Extra mixing due to planet-star interaction, like migration, could take place \citep{Castro}. The infall of planetary material might also affect the mixing processes of those stars \citep{theado} as well as the episodic accretion of planetary material \citep{Baraffe}. Finally a long-lasting star-disc interaction during pre-main sequence (PMS) evolution may cause the planet hosts to be slow rotators and develop a high degree of differential rotation between the radiative core and the convective envelope, also leading to enhanced lithium depletion \citep{Bouvier}, although the efficiency of rotational mixing is strongly reduced when the effects of magnetic fields are taken into account \citep{Eggenberger}.\\

Therefore, light elements are important tracers of stellar internal mixing and rotation. Since they are burned at relatively low temperatures we can obtain important information about the mixing and depletion processes that take place in the interior of these stars. At temperatures lower than 5600 K both stars with and without planets have their Li depleted because their convective envelopes are deep enough to reach the temperatures needed to burn Li. However, Be needs a greater temperature to be destroyed, so solar-type stars present shallow convective envelopes to reach the Be-burning layers. Therefore, if Be abundance is also affected by the presence of planets we would expect to see this effect in cooler stars.\\

In the literature there are many studies about Li in planet-host stars but this is not the case for the other light elements, Beryllium and Boron. B spectral lines are only visible from space so there are not studies of this element in stars with planets. On the other hand, measurements of Be abundances are difficulted by the fact that the only available lines are in the near-UV, a very line-blended region in metal-rich stars. Several studies show a wide dispersion in F and G field and cluster dwarf stars. In a recent work, \citet{Boesgaard09} found a wide range in Be abundances of more than a factor of 40 in spite of the similarity in the mass of the stars. Some of those stars, with very similar parameters, differed in Be abundance by a factor of 2. This spread in Be abundances might be explained by the different initial rotation rates of the stars. During the spin-down, extra mixing occurs leading to depletion of Li and Be. The maximum of Be content is found in the effective temperature range 5900-6300 K while hotter stars (6300-6500 K) fall in the Li-Be dip region where the depletion of those elements is atributted to slow mixing and is dependent upon age and temperature \citep{Boesgaard02,Boesgaard04}. \citet{Boesgaard04} found that Li was depleted more than Be in both temperature regimes and there was no difference in the depletion patterns between the cluster and field stars. \citet{Randich07} also showed that M67 members in the 5600-6100 T$_{\rm eff}$ range have not depleted any Be although they have depleted different amounts of Li.\\

\begin{figure}[ht!]
\centering
\includegraphics[width=8cm]{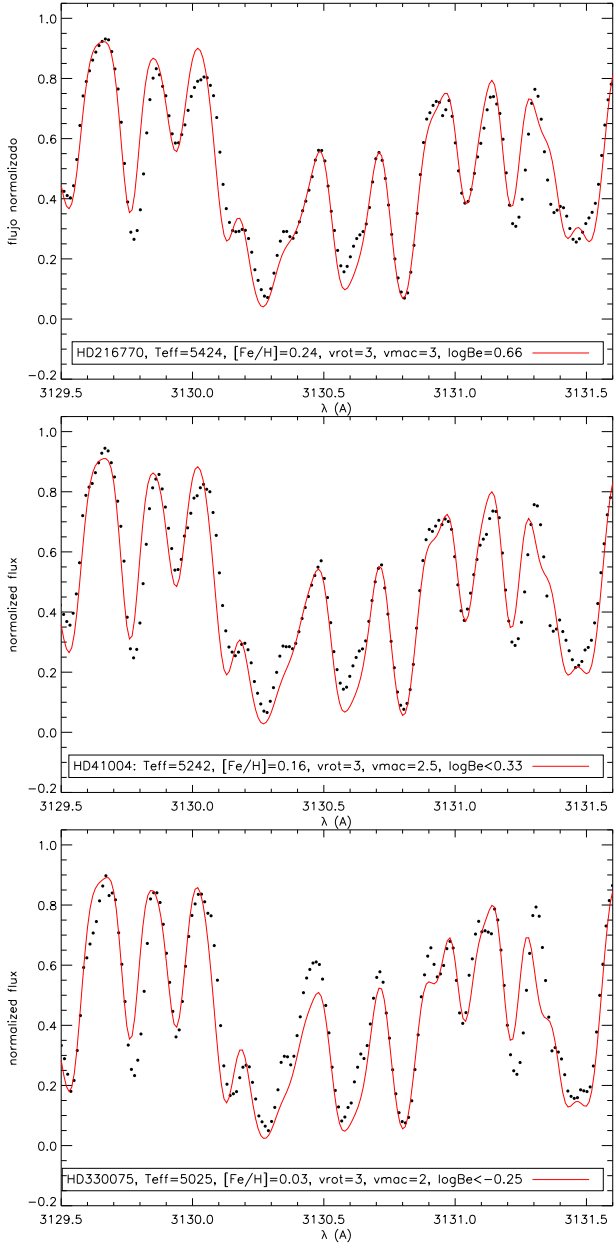}
\caption{Spectral syntheses (red lines) and observed spectra (dots) for the planet host stars HD 216770, HD 41004 and HD 330075.}
\label{be_syn}
\end{figure}

Some of the first works on Be abundances in stars with planets \citep{Garcia-Lopez,Deliyannis00} had a very small number of objects although \citet{Garcia-Lopez} already detected severe Be depletion in the planet host star 55 Cnc (T$_{\rm eff} \sim$ 5200 K). Our group has performed a extensive study of Be in the last years \citep{Garcia-Lopez,santos_be1,santos_be2,santos_be3,galvez}, analyzing a total sample of 70 and 30 stars with and without detected planets, respectively. Unfortunately, the number of cool dwarf stars in those samples is very small so they did not find any clear difference between the two populations. In this work we have measured Be abundances for 14 new stars, most of them with effective temperatures below 5600 K.

\section{Observations and analysis}
In this study we have obtained high resolution spectra for 14 new stars with magnitudes V between 6 and 10 using the UVES spectrograph at the 8.2-m Kueyen VLT (UT2) telescope (run ID 75.D-0453A) during the first semester of 2005. The dichroic mirror was used to obtain also red spectra and the slit width was 0.5 arcsec. These new spectra have a spectral resolution \textit{R} $\sim$ 70000 and \textit{S/N} ratios between 100 and 200. All the data were reduced using IRAF\footnote{IRAF is distributed by the National Optical Astronomy Observatories, operated by the Association of Universities for Research in Astronomy, Inc, under contract with the National Science Foundation, USA} tools in the \texttt{echelle} package. Standard background correction, flat-field and extraction procedures were used. The wavelength calibration was made using a ThAr lamp spectrum taken during the same night. Finally we normalized the spectra by fitting the observed spectra to a spline function of order 3 in the whole blue region (3040A-3800A).\\

The stellar atmospheric parameters were taken from \citet{santos04, santos05, sousa08}. The errors of the parameters from \citet{santos04, santos05} are of the order of 44 K for $T_{\rm eff}$, 0.11 dex for $\log g$, 0.08 km s$_{\rm -1}$ for $\xi_{t}$, 0.06 dex for metallicity and 0.05M$_{\odot}$ for the masses. From \citet{sousa08} the parameter errors are 25 K for $T_{\rm eff}$, 0.04 dex for $\log g$, 0.03 km s$_{\rm -1}$ for $\xi_{t}$ and 0.02 dex for metallicity. We refer to those works for further details in the parameters origin and errors. The three sets of parameters were determined in the same way although the most recent are more precise because they are calculated with more Fe lines. Thus, we will use these parameters when available. We note here the uniformity of the
adopted stellar parameters as discussed in Section 5 of \citet{sousa08}.\\

\begin{figure}[ht]
\centering
\includegraphics[width=8cm]{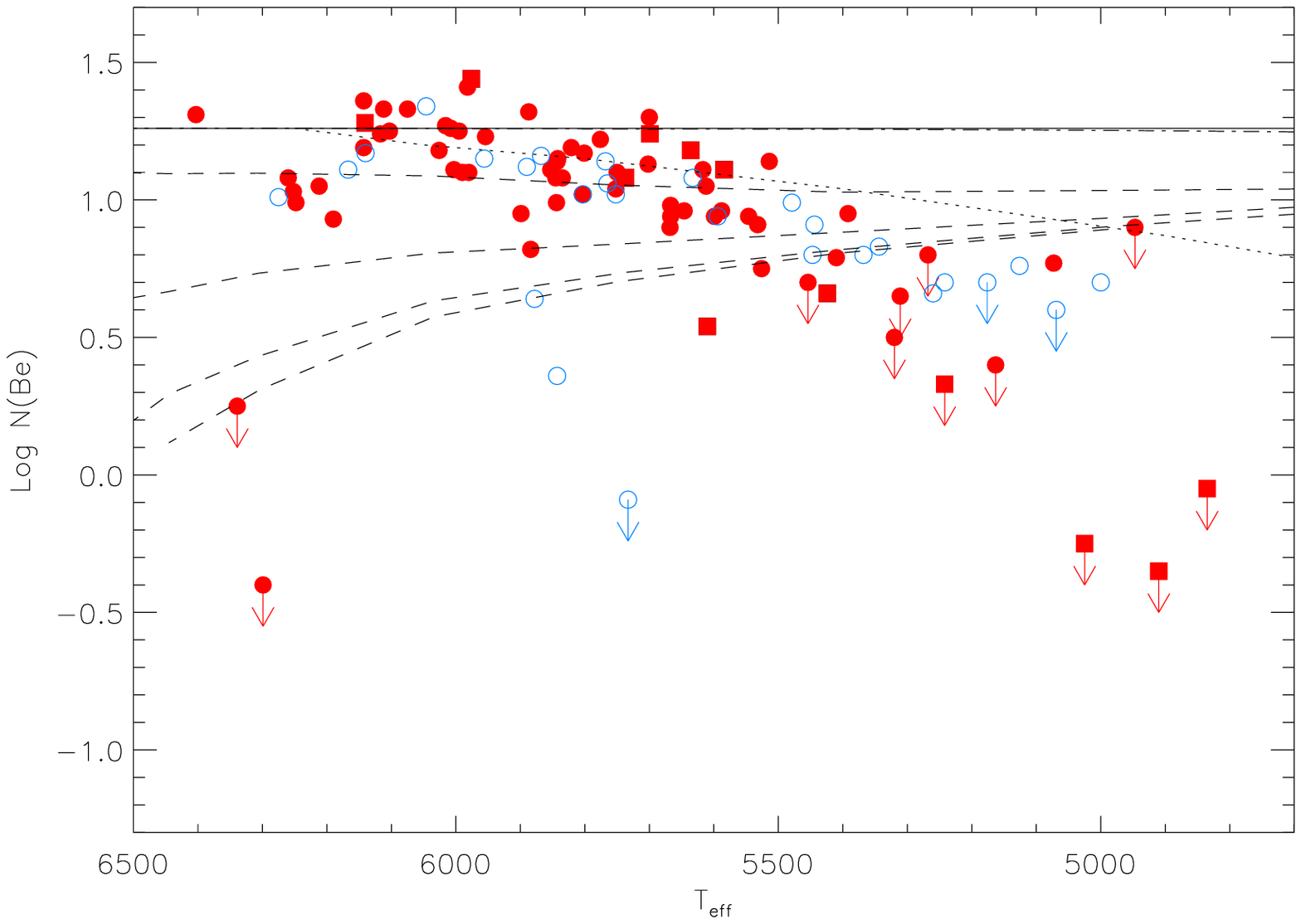}
\caption{Be abundances as a function of effective temperature for stars with planets from this work (red filled squares) and stars with (red filled circles) and without planets (blue open circles) from previous studies \citep{santos_be1,santos_be2,santos_be3,galvez}. The dashed lines represent 4 Be depletion models of \citet{Pinsonneault} (Case A) with different initial angular momentum for solar metallicity and an age of 1.7 Gyr. The solid line represents the initial Be abundance of 1.26. The dotted line represents the Be depletion isochrone for 4.6 Gyr taken from the models including mixing by internal waves of \citet{Montalban}.}
\label{be_teff}
\end{figure}

\begin{figure}[ht!]
\centering
\includegraphics[width=8cm]{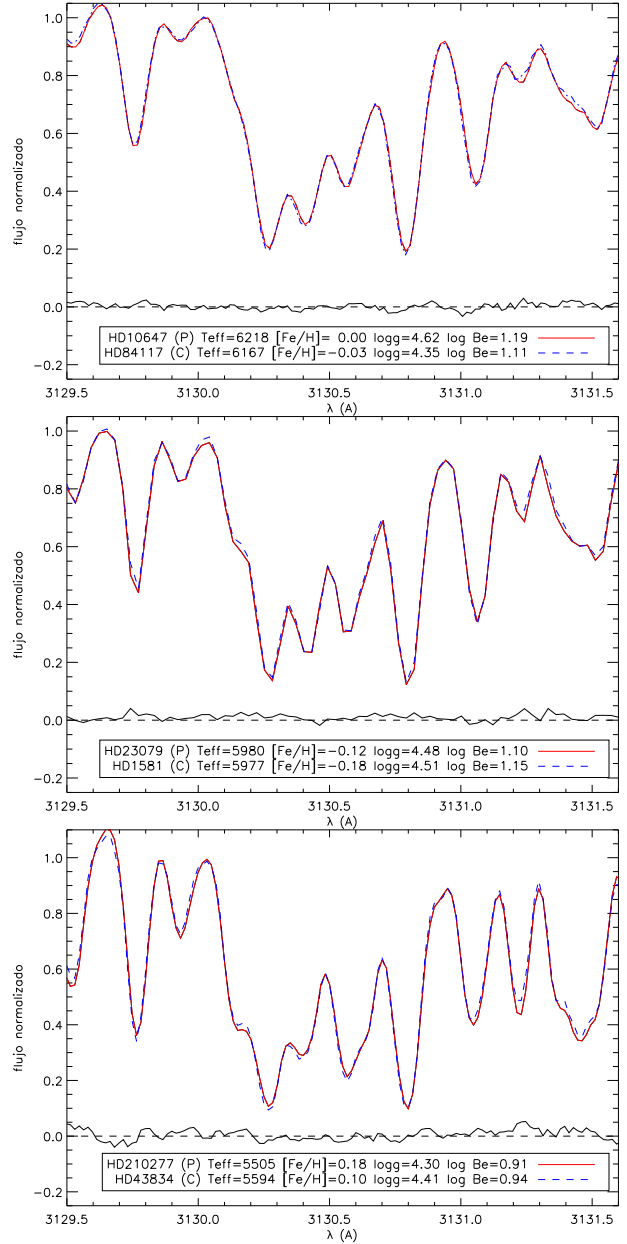}
\caption{Observed spectra and difference in fluxes for the planet-host stars HD 10647, HD 23079 and 210277 (red lines) and the comparison sample stars HD 84117, HD 1581 and HD 43834 (blue dashed lines). Stellar parameters of HD 10647, HD 23079, HD 1581 and HD 210277 are from \citet{sousa08}.}
\label{be_comp_calientes}
\end{figure}

Be abundances were derived by fitting the spectral region around the \ion{Be}{2} lines at 3130.420 and 3131.065 \AA{}, using the same line list as in \citet{Garcia-Lopez} although the bluest line was only used to check the consistency of the fit. These synthetic spectra were convolved with a rotational profile. We made a standard LTE analysis with the revised version of the spectral synthesis code MOOG2002 \citep{sneden} and a grid of Kurucz ATLAS9 atmospheres with overshooting \citep{kur93}. We give the total errors for Be abundances in Table \ref{tabla}. These errors are calculated adding quadratically the uncertainties due to the position of continuum and the errors in temperature, gravity and metallicity. Examples of shynthetic spectra are shown in Figure \ref{be_syn}, though we note that the results of this paper are not based on the Be abundance values obtained from spectral synthesis. Li abundances were taken from previous works (see Table \ref{tabla}) or measured using the same spectra as for Be abundances and the same spectral tools as in those works \citep{israelian04,israelian09}. Uncertainties for Li abundance determinations are between 0.05 and 0.1 dex.\\

\begin{deluxetable}{lccccrcr}
\tablecaption{Stars with planets analyzed in this work.\label{tabla}}
\tablewidth{0pt}
\tablehead{
\colhead{Star} & \colhead{T$_{\rm eff}$} & \colhead{log \textit{g}} & \colhead{$\xi_{t}$} & \colhead{[Fe/H]} & \colhead{log $\epsilon$(Be)} & \colhead{$\sigma$(Be)} & \colhead{log $\epsilon$(Li)}\\
\colhead{} & \colhead{[K]} & \colhead{[cm s$^{-2}$]} & \colhead{[km s$^{-1}$]}}
\startdata
HD 2039     &    5976   &   4.45   &   1.26  &    0.32   &     1.44 & 0.08 &     2.29\tablenotemark{2}\\ 
HD 4203     &    5636   &   4.23   &   1.12  &    0.40   &     1.18 & 0.10 & $<$0.70\tablenotemark{2}\\ 
HD 41004    &    5242   &   4.35   &   1.01  &    0.16   &  $<$0.33 & 0.16 & $<$0.44\tablenotemark{3}\\ 
HD 73526    &    5699   &   4.27   &   1.26  &    0.27   &     1.24 & 0.09 & $<$0.93\tablenotemark{3}\\ 
HD 76700    &    5737   &   4.25   &   1.18  &    0.41   &     1.08 & 0.10 & $<$1.39\tablenotemark{3}\\ 
HD 114386\tablenotemark{a}  &    4910   &   4.39   &   0.19  &   -0.07   & $<$-0.35 & 0.38 & $<$-0.14\tablenotemark{1}\\ 
HD 128311   &    4835   &   4.44   &   0.89  &    0.03   & $<$-0.05 & 0.15 & $<$-0.37\tablenotemark{2}\\ 
HD 154857   &    5610   &   4.02   &   1.30  &   -0.23   &     0.54 & 0.09 &     1.74\tablenotemark{3}\\ 
HD 177830\tablenotemark{b}   &    4804   &   3.57   &   1.14  &    0.33   & $<$-1.42 & --- & $<$-0.50\tablenotemark{2} \\ 
HD 190228\tablenotemark{b}   &    5327   &   3.90   &   1.11  &   -0.26   & $<$-0.09 & 0.18 & 1.23\tablenotemark{2}\\
HD 190360   &    5584   &   4.37   &   1.07  &    0.24   &     1.11 & 0.10 & $<$ 0.34\tablenotemark{2} \\ 
HD 208487\tablenotemark{a}   &    6146   &   4.48   &   1.24  &    0.08   &     1.28 & 0.07 &     2.75\tablenotemark{1} \\ 
HD 216770\tablenotemark{a}   &    5424   &   4.38   &   0.91  &    0.24   &     0.66 & 0.15 & $<$0.48\tablenotemark{1} \\ 
HD 330075\tablenotemark{a}   &    5025   &   4.32   &   0.63  &    0.03   & $<$-0.25 & 0.30 & $<$0.01\tablenotemark{1} \\ 

\enddata
\tablenotetext{a}{Parameters taken from Sousa et al. (2008)}
\tablenotetext{b}{Evolved stars}\\
\textbf{\tablenotetext{1}{Values from \citet{israelian09}}}\\
\textbf{\tablenotetext{2}{Values from \citet{israelian04}}}\\
\textbf{\tablenotetext{3}{Values from this work}}\\
\end{deluxetable}

\section{Discussion}
In Figure \ref{be_teff} we plot the derived Be abundances as a function of effective temperature for the planet-host stars in our sample (see Table \ref{tabla}) together with the previous samples. In this plot we have removed subgiants and giants to avoid evolutionary effects in the abundances. In this selection we took the spectral types of the stars from \citet{galvez}. Be abundances decrease from a maximum near T$_{\rm eff}$ = 6100 K towards higher and lower temperatures, in a similar way as Li abundances behave. In the high temperature domain, the steep decrease with increasing temperature resembles the well known Be gap for F stars \citep[e.g.][]{Boesgaard02}. The decrease of the Be content towards lower temperatures is smoother and may show evidence for continuous Be burning during the main sequence evolution of these stars. In the lowest temperature regime, around 5200 K, planet-host stars present lower abundances when compared with stars for which no planet has been discovered. Unfortunately, the number of comparison-sample stars in that temperature regime is still small. We note, however, the fact that only dwarf stars with planets show very low Be abundances at the lowest temperatures.\\

In addition, in this temperature range, we find very low abundances of Be for both groups of stars in contradiction with models of Be depletion \citep{Pinsonneault}. As already noticed in \citet{santos_be2,santos_be3}, these models agree with the observations above roughly 5600 K, but while the observed Be abundance decreases towards lower temperatures when T$_{\rm eff}$ $<$ 5600 K, these models predict either constant or increasing Be abundances. Even taking into account mixing by internal waves \citep{Montalban}, Be depletion is still lower than observed. Although uncertainties in Be abundandes for the coolest stars are large, Be content is still overestimated.\\

\begin{figure*}[ht]
\centering
\includegraphics[width=16cm]{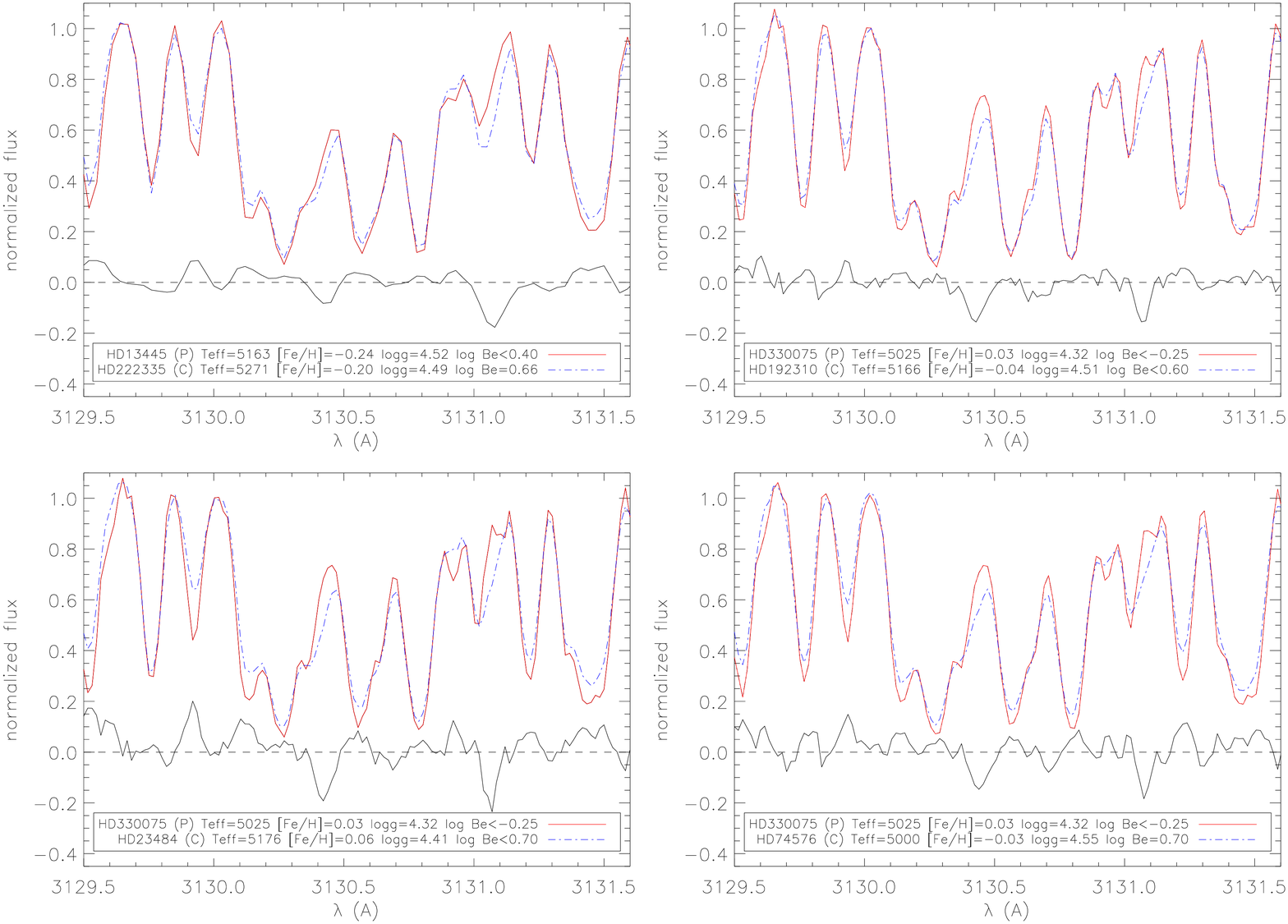}
\caption{Observed spectra and difference in fluxes for the planet-host stars HD 13445 and HD 330075 (red lines) and the comparison sample stars HD 222335, HD 192310, HD 23484 and HD 74576 (blue dashed lines).}
\label{be_comp}
\end{figure*}

To minimize the large uncertainties involved in the determination of Be abundances for cool stars we decided to make a direct comparison of pairs of spectra of similar stars and to carry out a differential Be abundance measurement. We searched for twin stars in the whole sample of 114 stars and found several pairs but only two cool planet host stars, HD 330075 and HD 13445 had analogs which we could properly use for a differential analysis. In Figure \ref{be_comp_calientes} we show the observed spectra for three pairs of stars with very similar parameters (T$_{\rm eff}$, log \textit{g} and [Fe/H]) and Be abundances. At those temperatures (T$_{\rm eff}$ $\gtrsim$ 5700K) a difference in the stellar parameters does not produce an important difference in Be abundance (see Table \ref{tabla_errores}), so if there are differences between the observed spectra, they cannot be due to effective temperature, gravity or metallicity effects. 
The difference in normalized flux between both stars of each pair (flux$_{\rm without ~planets}$ - flux$_{\rm planet ~host}$) is also plotted and, as we can see, it is small and practically constant along this narrow range of wavelengths. We evaluate this difference by calculating the standard deviation in the wavelength ranges 3130-3130.3 \AA{} and 3130.5-3130.85 \AA{} to avoid Be lines at 3130.420 and 3131.065 \AA{}. For these stars $\sigma$ = 0.008 and the difference in fluxes at the position of Be lines is at the level of 1$\sigma$. Subsequently, the slight difference in the Be content of these stars is due to the uncertainties, or in the case of the first couple, probably owing to the difference in gravity.\\

\begin{deluxetable}{lcccc}
\tablecaption{Uncertainties in Be abundances due to the variation in the stellar parameters using 4 stars with different effective temperatures as examples.\label{tabla_errores}}
\tablewidth{0pt}
\tablehead{
\colhead{Star} & \colhead{T$_{\rm eff}$} & \colhead{$\Delta$T$_{\rm eff}$ = $\pm$ 50K} & \colhead{$\Delta$log \textit{g} = $\pm$ 0.10} & \colhead{$\Delta$[Fe/H] = 0.05}}
\startdata
HD 208487    &  6141   &   $\mp$ 0.00   &  $\pm$   0.05   &  $\pm$ 0.02\\ 
HD 76700     &  5737   &   $\mp$ 0.02   &  $\pm$   0.07   &  $\pm$ 0.02\\
HD 216770    &  5424   &   $\mp$ 0.03   &  $\pm$   0.08   &  $\pm$ 0.02\\  
HD 330075    &  5025   &   $\mp$ 0.04   &  $\pm$   0.08   &  $\pm$ 0.02\\  
\enddata
\end{deluxetable}

It is worth mentioning that the second Be line is placed at the red wing of several Mn lines, the strongest one at 3131.037 \AA{}. For three out of four pairs of cool stars showed here, stars with planetary companions present always a higher Mn abundance than comparison sample stars \citep{gilli,neves}, hence this difference in fluxes is not due to Mn abundances.\\

\begin{deluxetable}{lccccccc}
\tablecaption{Analog stars analyzed in this work.\label{tabla_EW}}
\tabletypesize{\scriptsize}
\tablewidth{0pt}
\tablehead{
\colhead{Star} & \colhead{T$_{\rm eff}$} & \colhead{log \textit{g}} & \colhead{[Fe/H]} & \colhead{$\Delta$\textit{EW}$_{1}\rightarrow$ $\Delta$log(Be)$_{1}$}& \colhead{$\Delta$\textit{EW}$_{2}\rightarrow$ $\Delta$log(Be)$_{2}$} & \colhead{$\Delta$log(Be)} & \colhead{$\Delta$log(Be)$_{final}$}}
\startdata
HD 330075(P)   &    5025   &   4.32   &    0.03   & \nodata   & \nodata   & \nodata & \nodata \\ 
\noalign{\medskip}
\hline
\noalign{\medskip}
HD 192310(C)   &    5166   &   4.51  &   -0.04  &  12.7m\AA{} $\rightarrow$ 0.21   & 9.0m\AA{}$\rightarrow$ 0.28& 0.25 $\pm$ 0.05 & 0.21 $\pm$ 0.05\\ 
HD 74576(C)    &    5000   &   4.55  &   -0.03  &  13.4m\AA{}$\rightarrow$ 0.30   & 9.6m\AA{}$\rightarrow$ 0.33 & 0.32 $\pm$ 0.04 & 0.17 $\pm$ 0.04\\ 
HD 23484(C)    &    5176   &   4.41  &    0.06  &  16.3m\AA{}$\rightarrow$ 0.24   & 14.1m\AA{}$\rightarrow$ 0.42& 0.33 $\pm$ 0.10 & 0.32 $\pm$ 0.10\\ 
\noalign{\medskip}
\hline
\hline
\noalign{\medskip}
HD 13445(P)   &    5163  &   4.52   &  -0.24  &  \nodata  &  \nodata  & \nodata & \nodata\\ 
HD 222335(C)  &    5271  &   4.49   &  -0.20  &    8.5m\AA{}$\rightarrow$ 0.17  & 16.0m\AA{}$\rightarrow$ 0.42 & 0.29 $\pm$ 0.10 & 0.22 $\pm$ 0.10\\ 
\enddata
\tablecomments{The planet host stars HD 330075 and HD13445 are compared with three and one stars without detected planets, respectively. The difference in \textit{EW} between the lines of each pair are showed together with the abundance variation for these $\Delta$\textit{EWs}. Line 1 is at 3130.42 \AA{} and line 2 is at 3131.07 \AA{}. The final value of variation in Be abundance takes into account the effect of the differences between stellar parameters of each couple.}
\end{deluxetable}

\subsection{HD 330075}

In Figure \ref{be_comp} we display three comparison stars, HD 192310, HD 23484, HD 74576 with the same Be-depleted planet-host star, HD 330075. In the first pair (HD 330075 and HD 192310, top right panel) the standard deviation of the difference in fluxes has a value of $\sigma$ = 0.028, while the largest differences in that region are 0.156 at 3130.42 \AA{} and 0.155 at 3130.06 \AA{}. These differences are found exactly at the position of the two Be lines. Both values are at a level of 5$\sigma$, which confirms a measurable difference in Be lines. Although Mn abundances are slightly larger in stars with planets, the values are similar, so we consider that all stars have the same Mn abundance. Therefore, the difference in fluxes will correspond only to a difference in Be lines. We evaluate the difference in abundance by making a gaussian fit of the difference in fluxes. With this fit, centered at the position of Be lines, we estimate the equivalent width (\textit{EW}) of the difference in fluxes and then the difference in abundance associated to this $\Delta$\textit{EW} using MOOG2002 (\cite{sneden}). Since the two stars do not have exactly the same parameters, we calculate $\Delta$log(Be) using the models of both stars to see the differences. In this case, HD 192310 has a greater gravity than HD 330075, which gives a greater abundance (see Table \ref{tabla_errores}), but it also has a higher temperature which decreases the abundance, so the difference in abundance due to the stellar parameters is balanced and it is about 0.04 dex. The results are shown in Table \ref{tabla_EW}. For this pair of analog stars we find a final difference in Be abundances of about 0.21 dex.\\

For the second pair of stars, HD 330075 and HD 74576 (bottom right panel), the difference in flux has a standard deviation $\sigma$ = 0.037, while the biggest difference in flux is 0.169 and corresponds to 3131.07 \AA{}, at the position of the Be line. This difference in fluxes is again about 5$\sigma$. For these stars the difference in Be abundance found is about 0.32 dex. We note that for these two pairs of stars there is a good agreement between the difference in abundance yielded by the two lines. However, in this case, the comparison star has a similar temperature but a higher gravity, which produces a difference in abundance of 0.15 dex. Therefore, the initial difference in abundances (due to the difference in fluxes) is decreased by this quantity giving a final value of 0.17 dex.\\

The last comparison star for HD 330075, HD 23484 (bottom left panel), has higher T$_{\rm eff}$ and gravity, so a difference in Be abundance owing to the stellar parameters is balanced. In this case this difference is negligible, about 0.01 dex. We find again differences between the fluxes at the position of both Be lines of about 5$\sigma$. This gives an average final difference in abundance of 0.32 dex.\\

\subsection{HD 13445}

Finally, we study the pair HD 13445 and HD 222335 with similar parameters (see Figure \ref{be_comp}, top left panel). In this case, the difference in fluxes are at a level 7$\sigma$. However, this difference is in part due to the fact that the comparison star has a higher Mn abundance than the star with planets so we calculate the difference in \textit{EW} considering the Mn contribution. The variation of Mn abundance between both stars is 0.13 dex which is produced by a difference of 5 m\AA{} in the \textit{EW} of the MnI line, so we subtract this value to the measured \textit{EW}. We also estimate a variation of 0.07 dex in the abundance since the comparison star has a higher T$_{\rm eff}$ but a similar gravity. The final difference in abundances is about 0.22 dex. All these cool stars have depleted their Li (we only have upper limits for the abundances) so we cannot study the relation between Li and Be depletion.\\

Therefore, it seems that HD 330075 suffers an extra Be depletion when compared with analog stars without detected planets, and HD 13445 points to the same direction. Although spectral synthesis is difficult and the results present large uncertainties it is possible to measure differences in Be abundances in rather cool stars (T$_{\rm eff}\sim$ 5000 K) with analog stars.\\

\section{Conclusions}

We present new high-resolution UVES/VLT near-UV spectra of 14 stars with planets in order to find possible differences in Be abundances between those stars and a comparison sample of stars without detected planets. Be needs higher temperatures than Li to be destroyed so we have to search for these differences in cooler stars, which have deeper convective envelopes and are able to carry the material towards Be-burning layers. We find that stars with planets with T$_{\rm eff}$ $<$ 5500 K have on average Be abundances lower than comparison sample stars. Since uncertainties in Be abundances for cool stars are considerable, we decided to make a direct comparison between the observed spectra of twin stars with and without planets. In our sample, we only find analogs for two stars with planets in the lowest temperature regime. We find that for hot stars, with $T_{\rm eff} \gtrsim 5600$~K, both spectra are very similar but in the lowest temperature regime, the planet host star HD  330075 presents significant weaker lines of Be than three comparison stars without planets indicating an enhanced Be depletion by 0.25 dex. We analyze another planet host star, HD 13445, which suffers a similar depletion. Using this differential technique for twin cool stars it would be possible to investigate systematic differences in the Be depletion of planet host stars with respect to stars without planets in a temperature regime where systematic effects associated with the location of the continuum and poor knowledge of the atmospheric opacities may prevent an accurate absolute abundance analysis.\\

\acknowledgments
E.D.M, J.I.G.H. and G.I. would like to thank financial
support from the Spanish Ministry project MICINN AYA2008-04874.
J.I.G.H. acknowledges financial support from the Spanish Ministry 
project MICINN AYA2008-00695 and also from the Spanish Ministry of
Science and Innovation (MICINN) under the 2009 Juan de la Cierva
Programme.\\ 
N.C.S. would like to thank the support by the 
European Research Council/European Community under the FP7 through a 
Starting Grant, as well from Funda\c{c}\~ao para a Ci\^encia e a
Tecnologia (FCT), Portugal, through a Ci\^encia\,2007 
contract funded by FCT/MCTES (Portugal) and POPH/FSE (EC), 
and in the form of grant reference PTDC/CTE-AST/098528/2008
from FCT/MCTES.\\
This research has made use of the SIMBAD database
operated at CDS, Strasbourg, France. \\
This work has also made use of
the IRAF facility, and the Encyclopaedia of extrasolar planets.

\clearpage



\end{document}